\documentclass[conference]{IEEEtran}
\IEEEoverridecommandlockouts
\usepackage{cite}
\usepackage{amsmath,amssymb,amsfonts}
\usepackage{algorithmic}
\usepackage{graphicx}
\ifCLASSOPTIONcompsoc
\usepackage[caption=false,font=normalsize,labelfont=sf,textfont=sf]{subfig}
\else
\usepackage[caption=false,font=footnotesize]{subfig}
\fi
\usepackage{textcomp}
\usepackage{url}
\usepackage{xcolor}

\begin{document}

\bstctlcite{BSTcontrol}

\title{Investigating a Deep Learning Method to Analyze Images from Multiple Gamma-ray Telescopes
\thanks{This work was conducted in the context of the Analysis and Simulations Working Group of the CTA Consortium and makes use of simulated CTA data. We therefore acknowledge support from the agencies and organizations listed under Funding Agencies at this website: \protect\url{http://www.cta-observatory.org/}. AB acknowledges support for this work from by NSF award PHY-1229205. DN and TM acknowledge support from the Spanish MINECO / ERDF UE grant FPA2015-73913-JIN. We thank the NVIDIA GPU Grant Program for the donation of an NVIDIA Titan X GPU.}
}

\author{\IEEEauthorblockN{Aryeh Brill}
\IEEEauthorblockA{\textit{Department of Physics,} \\
\textit{Columbia University} \\
New York, New York \\
aryeh.brill@columbia.edu}
\and
\IEEEauthorblockN{Qi Feng}
\IEEEauthorblockA{\textit{Department of Physics} \\
\textit{and Astronomy, Barnard College}\\
New York, New York}
\and
\IEEEauthorblockN{T. Brian Humensky}
\IEEEauthorblockA{\textit{Department of Physics,} \\
\textit{Columbia University}\\
New York, New York}
\and
\IEEEauthorblockN{Bryan Kim}
\IEEEauthorblockA{\textit{Department of Physics}\\
\textit{and Astronomy, UCLA}\\
Los Angeles, California}
\and
\IEEEauthorblockN{Daniel Nieto}
\IEEEauthorblockA{\textit{Instituto de F\'{i}sica de Part\'{i}culas y del Cosmos,} \\
\textit{Universidad Complutense de Madrid}\\
Madrid, Spain}
\and
\IEEEauthorblockN{Tjark Miener}
\IEEEauthorblockA{\textit{Instituto de F\'{i}sica de Part\'{i}culas y del Cosmos,} \\
\textit{Universidad Complutense de Madrid}\\
Madrid, Spain}
}

\maketitle

\begin{abstract}
Imaging atmospheric Cherenkov telescope (IACT) arrays record images from air showers initiated by gamma rays entering the atmosphere, allowing astrophysical sources to be observed at very high energies. To maximize IACT sensitivity, gamma-ray showers must be efficiently distinguished from the dominant background of cosmic-ray showers using images from multiple telescopes. A combination of convolutional neural networks (CNNs) with a recurrent neural network (RNN) has been proposed to perform this task. Using CTLearn, an open source Python package using deep learning to analyze data from IACTs, with simulated data from the upcoming Cherenkov Telescope Array (CTA), we implement a CNN-RNN network and find no evidence that sorting telescope images by total amplitude improves background rejection performance.
\end{abstract}


\section{Motivation}

Very-high-energy (VHE; from about 20 GeV to 300 TeV) gamma rays provide a critical probe of the Universe's most extreme environments, offering the opportunity to study exotic astrophysics and fundamental physics at high energies and cosmological distances. Gamma rays in this energy range can be indirectly detected on the ground using arrays of imaging atmospheric Cherenkov telescopes (IACTs), which detect the Cherenkov light emitted from air showers produced by VHE gamma rays when they are absorbed by the atmosphere.

A wide variety of scientific studies can be performed with VHE gamma rays \cite{Consortium2019}. VHE gamma rays are observed from supernova remnants and pulsar wind nebulae in the Milky Way and supermassive black holes in distant galaxies, providing insight into the nature of these sources, such as how and where in these sources particles are accelerated to relativistic energies. Astrophysicists also search for VHE gamma-ray emission from dark-matter-dominated objects such as dwarf galaxies, looking for gamma rays hypothesized to be produced by dark matter annihilation or decay. In addition, IACTs play a key role in multimessenger astronomy, regularly searching for VHE emission produced by gamma-ray bursts and by the sources of gravitational wave events, and having recently detected TeV gamma-ray emission from a flaring blazar coincident with a highly energetic neutrino detected by the IceCube Neutrino Observatory \cite{Aartsen2018}.

\begin{figure*}[!t]
\centering
\subfloat[Example IACT image]{\includegraphics[width=0.85\columnwidth]{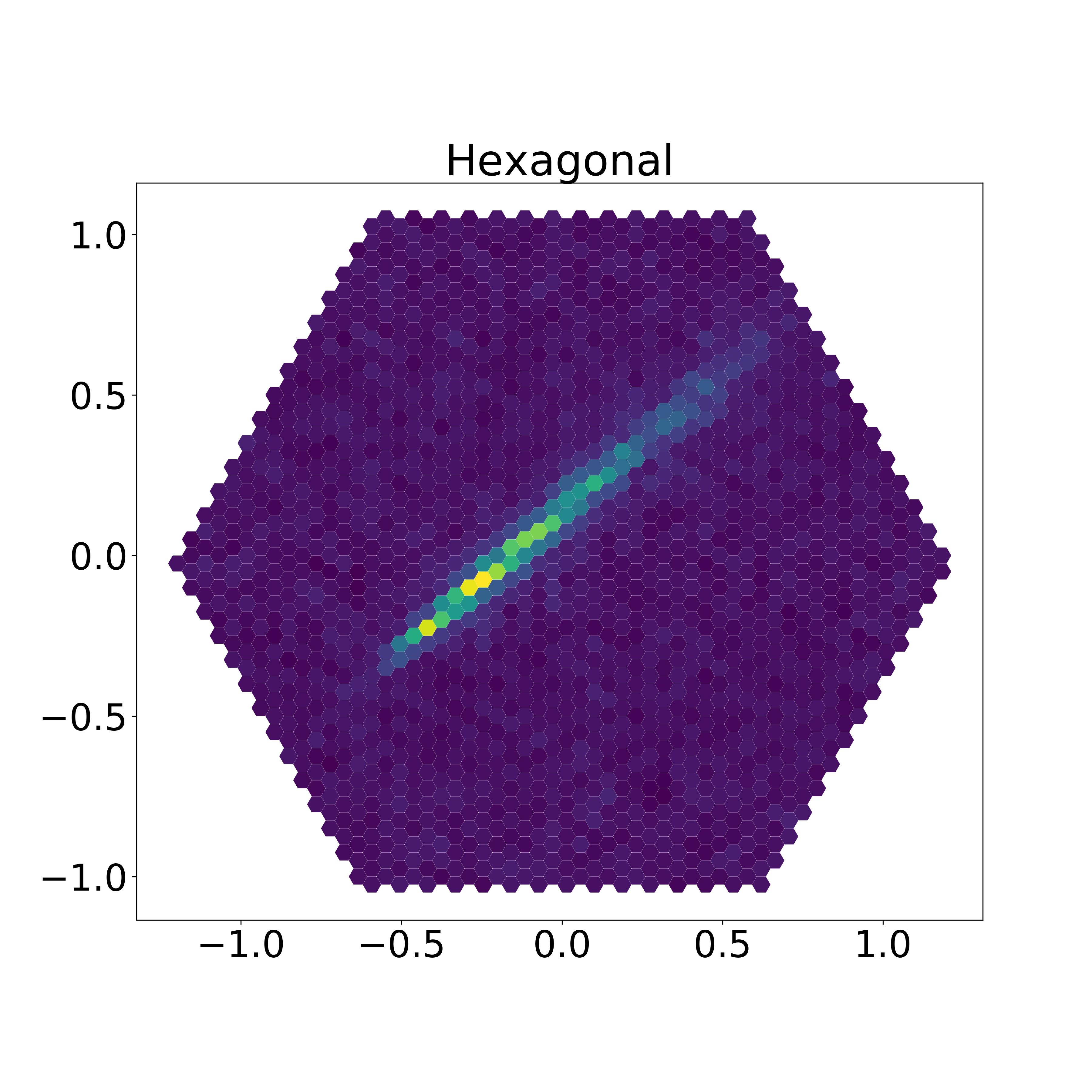}
\label{fig:flashcam}}
\hfil
\subfloat[Same image mapped using rebinning]{\includegraphics[width=0.85\columnwidth]{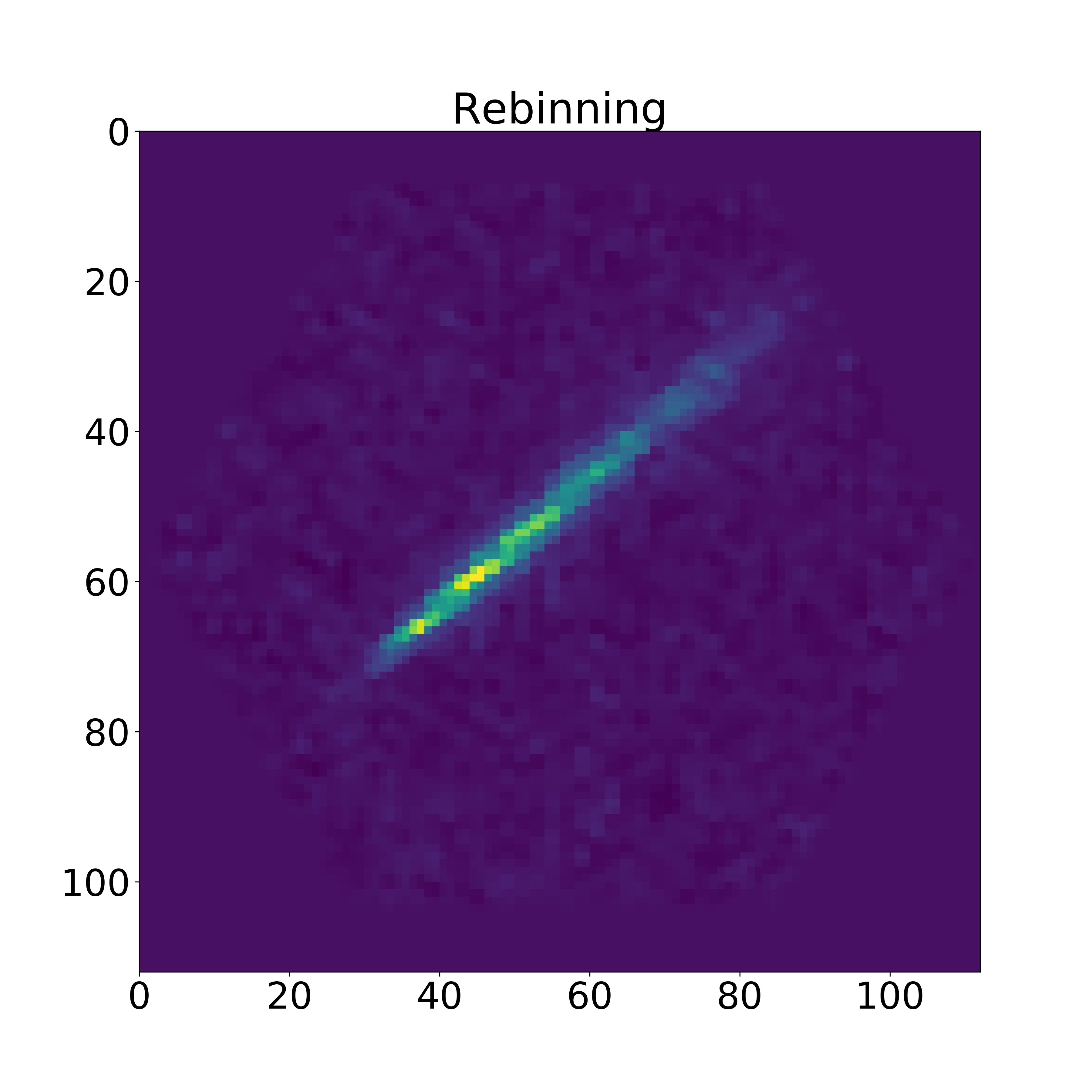}
\label{fig:rebinning}}
\caption{Left: An example IACT image from a CTA FlashCam camera simulation, illustrating the hexagonally spaced grid of pixels typical of many IACT cameras. Right: The same image mapped to a square matrix of pixels by rebinning, which preserves the image's overall amplitude. Both images are from \cite{Nieto2019b}.}
\label{fig:mapping_methods}
\end{figure*}

Measurements with IACTs enable these scientific studies by extracting information about VHE particles from the air showers they produce in the atmosphere. In a conventional IACT analysis, images from multiple telescopes are parameterized and stereoscopically combined to extract the spatial, temporal, and calorimetric information of the originating VHE particle.

\section{Gamma-ray Image Analysis}

The sensitivity of IACTs depends strongly on efficiently rejecting the background of much more numerous cosmic-ray showers, which resemble those produced by gamma rays but tend to have a more complex morphology. Using the information contained in the shapes of the shower images is therefore critical to maximizing IACT sensitivity. Supervised learning algorithms, like random forests and boosted decision trees, have been shown to effectively classify IACT events based on event-level parameters constructed using images from multiple telescopes (e.g. \cite{Krause2017}).

Deep learning techniques, such as convolutional neural networks (CNNs), may be used to improve on these methods because they do not require the images to be parameterized and may therefore access features of these images that would be washed out by the parameterization \cite{Nieto2017}. A deep learning approach that combines CNNs with a recurrent neural network (RNN) has been shown to improve background rejection performance using data from the H.E.S.S. IACT array \cite{Shilon2019}. In previous work, the input images to such a network have been sorted by total amplitude. In this study, we apply a similar model to simulated data from the Cherenkov Telescope Array (CTA) \cite{Acharya2013}, the next-generation observatory for gamma-ray astronomy, to determine the effect of this sorting procedure on classification performance. 

\section{CTLearn}

We implement our neural network model using CTLearn\footnote{\url{https://github.com/ctlearn-project/ctlearn}} \cite{ari_brill_2019_3345947}, an open-source Python package for using deep learning to analyze pixel-wise camera data from arrays of IACTs. CTLearn provides an application-specific framework for configuring and training machine learning models with TensorFlow\footnote{\url{https://www.tensorflow.org}} and applying the trained models to generate predictions on a test set \cite{Nieto2019a}. CTLearn v0.3.0 was used for training the models used in this work.

Through the associated DL1-Data-Handler package \cite{bryan_kim_2019_3336561}, CTLearn can load and preprocess IACT data from any major current- or next-generation IACT. In particular, because many IACT cameras have pixels arranged in a hexagonal layout, posing a challenge for convolutional neural networks that conventionally require as input a rectangular matrix of input pixels, DL1-Data-Handler provides a number of methods to map hexagonally spaced pixels to a square grid. In this work, the rebinning method was chosen (Fig. \ref{fig:rebinning}), which is one of several mapping methods that provide comparably good performance \cite{Nieto2019b}.

\begin{figure}[htbp]
\centerline{\includegraphics[width=0.85\columnwidth]{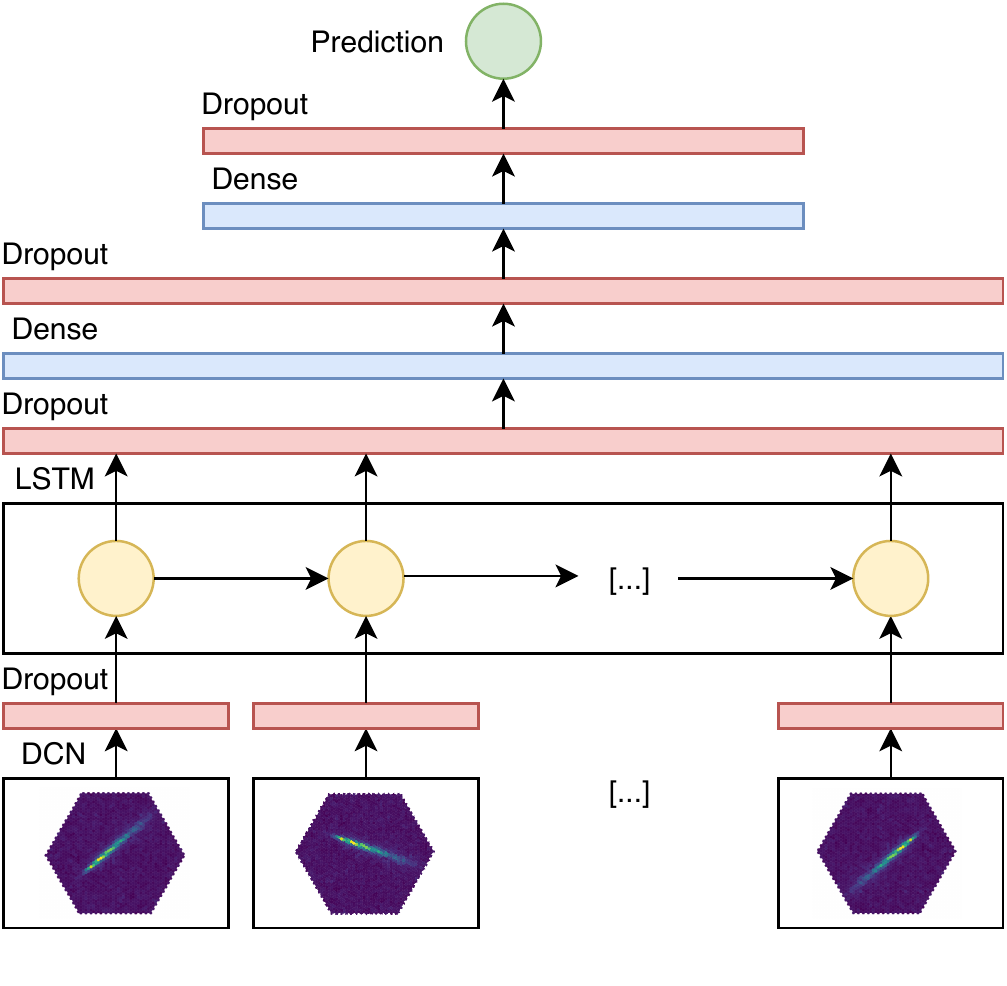}}
\caption{Diagram of the CNN-RNN particle classification model implemented in CTLearn, from \cite{Nieto2019a}. The model uses a CNN block (labeled as a deep convolutional network or DCN) to derive a vector representation of each image in an event. The vectors are combined using a Long Short Term Memory network (LSTM), a type of recurrent neural network (RNN).}
\label{fig:cnn_rnn}
\end{figure}

\section{CNN-RNN Particle Classification Model}

A challenge of using deep learning methods with IACT data is combining images from multiple telescopes providing different views of an air shower event. Each event triggers multiple telescopes, and the number of triggered telescopes may vary from event to event.

One approach to deal with this challenge is to break the problem into two stages. First, each image is processed into a vector representation by a CNN, using the same weight parameters for each image. The vectors are then combined by a recurrent neural network (RNN), a type of neural network that takes as input a sequence of vectors, and, by maintaining an internal state, produces an output vector that depends not only on the most recent input but on all preceding inputs in the sequence. This vector is then fed into a set of densely connected layers that produce the final prediction. Connecting these networks allows a single model trained end-to-end to classify events consisting of images from multiple telescopes.

For this work, the built-in CNN-RNN model of CTLearn was used, which implements an architecture similar to the CRNN network presented in \cite{Shilon2019}. More details on the model and the default hyperparameter settings that were used can be found in \cite{Nieto2019a}. The RNN in this model is specifically a Long Short-Term Memory (LSTM) network.

Recurrent neural networks are capable of processing sequential data in which the ordering of inputs may affect their interpretation. Therefore, having a meaningful ordering of telescope images in a CNN-RNN network may improve performance. In previous work using a CNN-RNN network for classifying Cherenkov air showers as produced by a gamma ray or a cosmic-ray proton, the telescope images were ordered by total image amplitude, or size. As size can be considered to be a proxy for proximity to the shower center, sorting on this parameter may provide an ordering given the absence of temporal information \cite{Shilon2019}.

To understand the effect of this ordering on performance, we trained two CNN-RNN networks as described above to classify IACT images as produced by a gamma ray or a cosmic-ray proton, changing only the ordering of the input images. As a control, in one network the images were ordered by telescope ID number, an arbitrary but consistent ordering, while in the other the images were ordered by size. The networks were trained using a sample of 250,000 simulated events from 25 FlashCam telescopes \cite{Gadola2015}, part of a proposed CTA array in Paranal, Chile. Ten percent of the events in the sample were reserved as a validation set, which was not used for training.

\begin{figure}[htbp]
\centerline{\includegraphics[width=0.85\columnwidth]{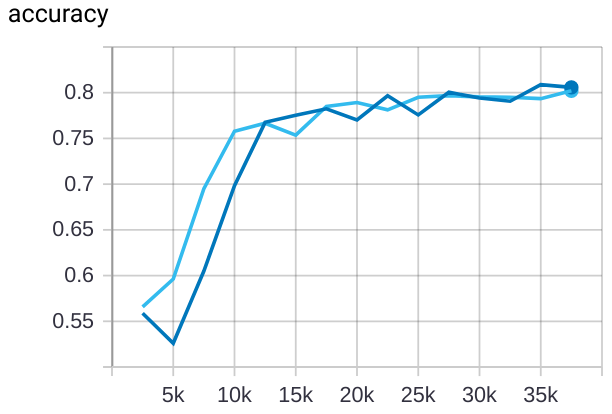}}
\caption{Validation accuracy of the CNN-RNN model with images ordered by ID (dark blue) and total brightness (light blue) as a function of number of training steps (batches of 16 events). The models reach respective accuracies of 80.6\% and 80.2\%.}
\label{fig:accuracy}
\end{figure}

\begin{figure}[htbp]
\centerline{\includegraphics[width=0.85\columnwidth]{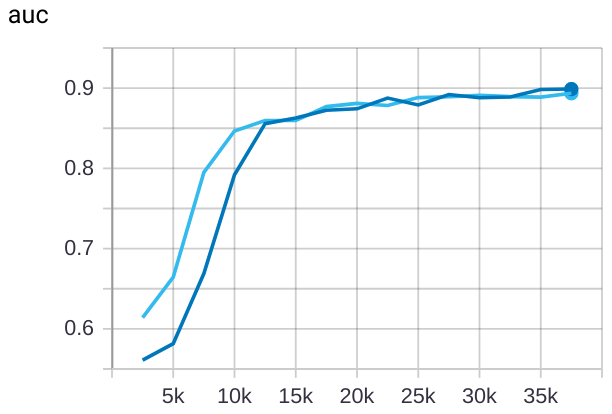}}
\caption{Validation AUC with images ordered by ID (dark blue) and total brightness (light blue) as a function of number of training steps (batches of 16 events). AUC is the numerically integrated area under the receiver operating characteristic curve, measuring sensitivity and specificity. The models reach respective AUCs of 0.899 and 0.894.}
\label{fig:auc}
\end{figure}

\section{Results and Discussion}

The results of this experiment are shown in Fig.~\ref{fig:accuracy} and Fig.~\ref{fig:auc}. The validation metrics of the two models were approximately the same, with those of the control model being slightly higher. The control model attained validation accuracy and AUC of 80.6\% and 0.899, while the model with images sorted by size reached 80.2\% and 0.894. We therefore find no evidence that sorting images by size improves gamma-proton classification performance with a CNN-RNN model.

This finding leaves open the possibility that a different ordering of telescope images could result in improved performance. In particular, an ordering which provides sufficient information about the telescopes' position on the ground could help a CNN-RNN to perform stereoscopic reconstruction of Cherenkov air showers. While ordering by size as a proxy for distance to the shower center should provide some relative position information, it is possible this information is too incomplete to be useful to the network. 

In addition to performing background rejection, deep learning algorithms could be used to determine the arrival direction and energy of the particles initiating Cherenkov air showers \cite{Mangano2018}, tasks for which stereoscopic reconstruction is particularly important. Ensuring that telescope position information is effectively provided to CNN-RNN networks may therefore not only improve their performance on background rejection but also on additional tasks critical for IACT image analysis.

\bibliographystyle{IEEEtran}
\bibliography{IEEEabrv,main.bib}

\end{document}